\documentclass[aps,pra,reprint,superscriptaddress,showpacs]{revtex4-1}
\usepackage{graphicx}
\usepackage{amsmath}
\usepackage{textcomp}
\usepackage{color}
\usepackage{xcolor}
\usepackage{upgreek}
\usepackage{comment}
\usepackage{verbatim}
\usepackage{amssymb}
\usepackage{tabu}
\usepackage{float}
\usepackage[english]{babel}
\usepackage{gensymb}
\usepackage{mathtools}
\begin{document}
\setlength{\abovedisplayskip}{5pt}
\setlength{\belowdisplayskip}{5pt}
\title{Shaping Field Gradients for Nanolocalization}



\author{Sergey Nechayev}
\affiliation{Max Planck Institute for the Science of Light, Staudtstr. 2, D-91058 Erlangen, Germany}
\affiliation{Institute of Optics, Information and Photonics, University Erlangen-Nuremberg, Staudtstr. 7/B2, D-91058 Erlangen, Germany}

\author{J{\"o}rg S.~Eismann}
\affiliation{Max Planck Institute for the Science of Light, Staudtstr. 2, D-91058 Erlangen, Germany}
\affiliation{Institute of Optics, Information and Photonics, University Erlangen-Nuremberg, Staudtstr. 7/B2, D-91058 Erlangen, Germany}
\author{Martin Neugebauer}
\affiliation{Max Planck Institute for the Science of Light, Staudtstr. 2, D-91058 Erlangen, Germany}
\affiliation{Institute of Optics, Information and Photonics, University Erlangen-Nuremberg, Staudtstr. 7/B2, D-91058 Erlangen, Germany}
\author{Peter Banzer}
\email[]{peter.banzer@mpl.mpg.de}
\homepage[]{http://www.mpl.mpg.de/}
\affiliation{Max Planck Institute for the Science of Light, Staudtstr. 2, D-91058 Erlangen, Germany}
\affiliation{Institute of Optics, Information and Photonics, University Erlangen-Nuremberg, Staudtstr. 7/B2, D-91058 Erlangen, Germany}
\date{\today}
\begin{abstract}
Deep sub-wavelength localization and displacement sensing of optical nanoantennas have emerged as extensively pursued objectives in nanometrology, where focused beams serve as high-precision optical rulers while the scattered light provides an optical readout. Here, we show that in these schemes using an optical excitation as a position gauge implies that the sensitivity to displacements of a nanoantenna depends on the spatial gradients of the excitation field. Specifically, we explore one of such novel localization schemes based on appearance of transversely spinning fields in strongly confined optical beams, resulting in far-field segmentation of left- and right-hand circular polarizations of the scattered light, an effect known as the giant spin-Hall effect of light. We construct vector beams with augmented transverse spin density gradient in the focal plane and experimentally confirm enhanced sensitivity of the far-field spin-segmentation to lateral displacements of an electric-dipole nanoantenna. We conclude that sculpturing of electromagnetic field gradients and intelligent design of scatterers pave the way towards future improvements in displacement sensitivity.
\end{abstract}
\maketitle
\section{Introduction}
\begin{figure}
  \includegraphics[width=0.48\textwidth]{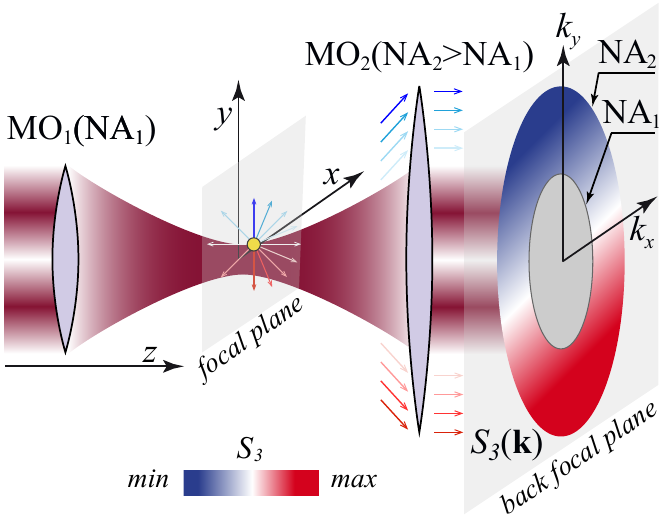}
  \caption{Spin-segmentation-based displacement detection scheme. The first microscope objective (MO$_1$) focuses the beam onto a nanoantenna. Another confocally aligned microscope objective (MO$_2$) with larger numerical aperture (NA) collects the transmitted and the scattered light. We analyze the circular polarization distribution of the scattered light in the back focal plane ($k$-space) of MO$_2$. The degree of circular polarization is expressed by the angularly resolved third Stokes parameter $S_3(\mathbf{k})$. Owing to the larger NA of MO$_2$, the angular region of NA$>$NA$_1$ contains only the scattered light.}
  \label{fig:sketch}
\end{figure}

Currently, there is a growing interest in localization and displacement sensing of optical nanoantennas~\cite{bharadwaj_optical_2009,novotny_antennas_2011,curto_multipolar_2013,krasnok_optical_2013} in focal volumes of tightly focused beams. Tight focusing produces highly inhomogeneous three-dimentional (3D) focal fields~\cite{novotny_principles_2012}, resulting in position-dependent excitation and scattering of nanoantennas~\cite{Nugent-Glandorf_Measuring_2004,rodriguez-herrera_far-field_2010,rodriguez-herrera_optical_2010,neugebauer_polarization_2014,roy_radially_2015,xi_accurate_2016,neugebauer_polarization-controlled_2016,xi_magnetic_2017,xi_retrieving_2018,Bag2018,Lorenz-Mie_non_parax_2018,Wei_Interferometric_2018,Neugebauer_Weak_2019}. In this schemes, information on the position of a scatterer with respect to the focal fields becomes encoded in the directivity and polarization pattern of the far-field scattered light. The latter implies that the far-field sensitivity to displacements of a nanoantenna may be enhanced by tailoring the spatial derivatives of the excitation field.\\
In this letter, we consider far-field spin-segmentation of the scattered light --- an effect referred to as the giant spin-Hall effect of light~\cite{rodriguez-herrera_optical_2010,rodriguez-herrera_far-field_2010,Nechayev_Weak_2018,Neugebauer_Weak_2019,Olmos-Trigo:19} --- as a measure of the position of an isotropic electric-dipole nanoantenna located in the focal plane of a tightly focused beam. The extent of this far-field circular polarization splitting depends on the eccentricity and orientation of the polarization ellipse of the excited electric dipole moment. 
The ``maximal'' spin splitting is achieved for a dipole moment whose vector upon time evolution draws a circle lying in the propagation (meridional) plane with respect to the incident light~\cite{Nechayev_Weak_2018}. 
Consequently the extent of the spin segmentation depends on the transverse components of the spin density vector at the position of the nanoantenna~\cite{aiello_transverse_2009,banzer_photonic_2013,bekshaev_transverse_2015,bliokh_transverse_2015,aiello_transverse_2015,neugebauer_magnetic_2017}. To enhance the sensitivity of the far-field spin segmentation to displacements of the scatterer, we experimentally utilize a generalized cylindrical vector beam~\cite{spiral_beam,mimicking2019} with augmented transverse spin density (TSD) gradient. Our study based on structuring the gradient of the focal TSD reveals that the displacement sensitivity can be drastically increased. Thus, our work may constitute an important step in nanophotonics and localization microscopy.\\
%
%
A schematic of the spin-segmentation-based displacement detection scheme is shown in Fig.~\ref{fig:sketch}. A polarization tailored beam is focused by a microscope objective (MO$_1$) with the numerical aperture NA$_1$ onto a nanoantenna positioned in the focal plane. Highly inhomogeneous 3D focal fields result in position dependent scattering patterns. We collect the transmitted and the scattered light by a second microscope objective (MO$_2$) with a higher numerical aperture NA$_2$, creating an angular region in the back focal plane (BFP) of MO$_2$, which allows us to analyze the far-field scattered light only. This is achieved by blocking the angular region containing the transmitted excitation beam. We detect the changes in the far-field distribution of circular polarization, expressed by the angularly resolved third Stokes parameter $S_3(\mathbf{k}$), and link these changes to the displacements of the nanoantenna in the focal plane.%
\\
\begin{figure}
  \includegraphics[width=0.48\textwidth]{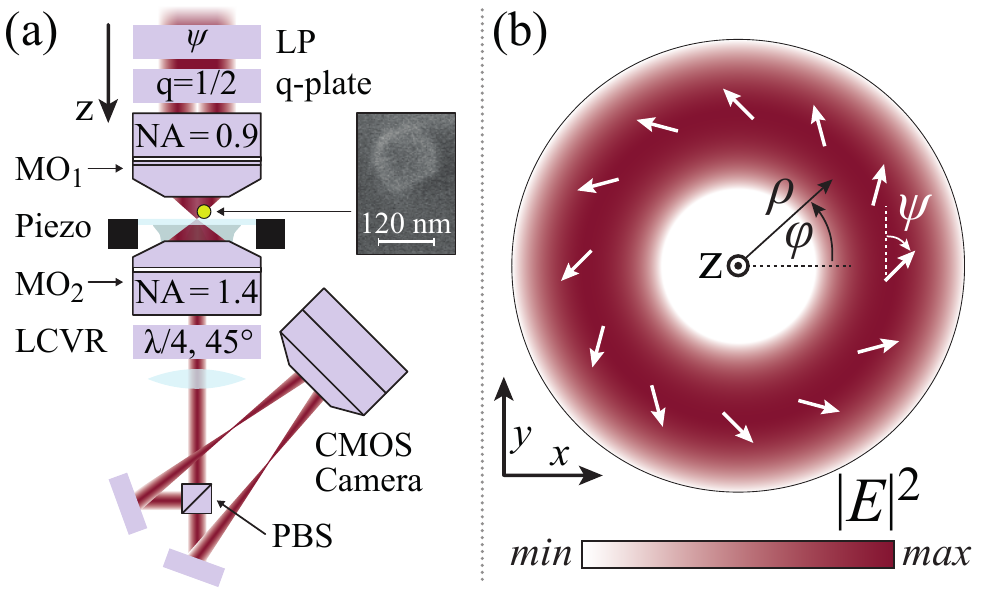}
  \caption{$(\boldsymbol{a})$ Sketch of the experimental setup. A linear polarizer (LP), a q-plate and a Fourier filter (not shown) prepare the incident beam. 
	The beam is focused by an aplanatic dry microscope objective (MO$_1$) and excites a gold nanoparticle (see scanning-electron microgaph shown as inset) of diameter $d = 120\,$nm. A confocally aligned index-matched immersion microscope objective (MO$_2$) collects the transmitted and scattered light. We split the outgoing beam into left- and right-hand circularly polarized components using a liquid crystal variable retarder (LCVR) and a polarizing beam splitter (PBS) and simultaneously image them onto a CMOS camera. $(\boldsymbol{b})$ Electric field distribution $\mathbf{E}_{\text{in}}$ at the entrance pupil of MO$_1$. The intensity pattern $|\mathbf{E}_{\text{in}}|^2$ corresponds to the colormap. The angle $\psi$ defines the polarization pattern, which is indicated by white arrows.}
  \label{fig:setup}
\end{figure}%
\section{Theory}
First, we assume that MO$_1$ focuses an incident paraxial beam, propagating along the $z$-axis, onto a spherical isotropic point-dipole nanoantenna (compare Fig.~\ref{fig:sketch} and Fig.~\ref{fig:setup}$\boldsymbol{a}$). The focal electric $\mathbf{E}^\text{foc}\left(x,y\right)$ and magnetic $\mathbf{H}^\text{foc}\left(x,y\right)$ fields  excite position-dependent electric $\mathbf{p}\left(x,y\right)=\alpha_e \varepsilon_0 \mathbf{E}^\text{foc}\left(x,y\right)$ and magnetic $\mathbf{m}\left(x,y\right) = \alpha_m \mathbf{H}^\text{foc}\left(x,y\right)$ dipole moments in the nanoantenna~\cite{novotny_principles_2012,jackson_classical_1999}. Here, $\alpha_e$ and $\alpha_m$ are the electric-dipole and magnetic-dipole polarizabilities~\cite{Bohren_Absorption_1983}, respectively, and $\varepsilon_0$ is the vacuum permittivity. The free space far-field emission $\mathbf{E}^{f}$ of this point-dipole (see Appendix A) in transverse $x$-direction $\mathbf{k}=(\pm k,0,0)$, where $k$ is the vacuum wavenumber, is given by:
\begin{align}
\begin{split}
\label{eqn:approxfarfield}
 &\mathbf{E}^{f}(\pm k,0,0)=
\left[\begin{matrix}
\mathrm{E}^{f}_{p}(\pm k)\\
\mathrm{E}^{f}_{s}(\pm k)\\
\end{matrix}\right]
\propto
\left[\begin{matrix}
-p_z \pm m_y/c \\
\pm p_y + m_z/c
\end{matrix}\right] \left(x,y\right)
 \text{,}\\
\end{split}
\end{align}
where the subscripts $p$ and $s$ indicate the transverse magnetic (TM) and electric (TE) field components, respectively, and $c$ is the speed of light in vacuum. We omit the spatial dependence in most of the following equations for clarity. Using Eqn.~\ref{eqn:approxfarfield}, we define the far-field spin-segmentation $D_x$ with respect to $k_x$ as~\cite{Neugebauer_Weak_2019}:
\begin{align}
D_x\left[  \mathbf{p}\left(x,y\right),\mathbf{m}\left(x,y\right)\right] =\frac{S_3\left(+k\right) -S_3\left(-k\right) }{S_0\left(+k\right) +S_0\left(-k\right)}
\label{eqn:dir1} 
\text{,}
\end{align}
where $S_3=-2 \Im \left\{  \mathrm{E}^{f}_{p}  \left( \mathrm{E}^{f}_{s} \right)^*  \right\}$ is the third Stokes parameter describing the preponderance of left- over right-hand circularly polarized light (LCP and RCP) and $S_0=| \mathrm{E}^{f}_{p}|^2 +  |\mathrm{E}^{f}_{s} |^2$ describes the far-field intensity distribution. Substituting expressions for the dipole moments in Eqn.~\ref{eqn:dir1} relates $D_x$ to the focal fields:
\begin{align}
D_x&=  \frac{  2  \Im  \left\{  \mathrm{E}_z^\text{foc} \left( \mathrm{E}_y^\text{foc}  \right) ^*  +\eta^2 \beta  \mathrm{H}_z^\text{foc} \left( \mathrm{H}_y^\text{foc}  \right) ^*  \right\}    }{   \left( |\mathrm{E}_z^\text{foc} |^2+| \mathrm{E}_y^\text{foc} |^2 \right) + \eta^2 \beta \left( |\mathrm{H}_z^\text{foc} |^2+| \mathrm{H}_y^\text{foc} |^2 \right) }
\label{eqn:dir2_full} 
\text{,}
\end{align}
where $\eta$ is the vacuum impedance and $\beta \equiv |\alpha_m / \alpha_e|^2$ is the squared ratio of the dipole polarizabilities. For an electric-dipole $(\beta = 0)$ nanoantenna, $D_x(\beta = 0)$ depends on the normalized $x$-component of electric TSD of the focal fields $\mathbf{s}_\text{E}  \propto  \Im \left\{\mathbf{E}^{\text{foc}\,\ast} \times \mathbf{E}^\text{foc}\right\}$~\cite{aiello_transverse_2009,banzer_photonic_2013,bekshaev_transverse_2015,bliokh_transverse_2015,aiello_transverse_2015,neugebauer_magnetic_2017}:
\begin{align}
D_x(\beta = 0)  =  \frac{ 2   \Im  \left\{ \mathrm{E}_z^\text{foc} \left( \mathrm{E}_y^\text{foc}  \right) ^*  \right\}    }{| \mathrm{E}_y^\text{foc} |^2+| \mathrm{E}_z^\text{foc} |^2 }
\label{eqn:dir2} 
\text{.}
\end{align}
The sensitivity to displacements of the scatterer across the focal plane $\Upsilon_x$ and the angle $\theta_x$ of the steepest ascent of $D_x$ with respect to the $x$-axis are:  
\begin{align}
\begin{split}
&\Upsilon_x= |\nabla D_x|,\\
&\theta_x = \tan^{-1}   \left( \partial_y D_x , \partial_x D_x     \right)
\text{,}
\end{split}
\label{eqn:sens1}
\end{align} 
where $\tan^{-1}$ is the four-quadrant inverse tangent. As mentioned before, our main goal is to enhance the sensitivity of the spin-segmentation in order to measure small particle displacements. Therefore, to enhance $\Upsilon_x$, we aim to design a field with an electric TSD distribution exhibiting the steepest possible gradient in the focal plane. As an example of such a field, we consider an incident spirally-polarized cylindrical vector beam~\cite{spiral_beam,mimicking2019}, shown in Fig.~\ref{fig:setup}$\boldsymbol{b}$, which is a linear superposition of in-phase radially and azimuthally polarized beams:
\begin{align}
\begin{split}
\mathbf{E}_{\text{in}} =E_0\frac{\rho}{w_0}e^{-\frac{\rho^2}{w_0^2}}  (\, \underbrace{  \sin \psi \hat{\boldsymbol {\rho}}}_{\text{Radial}} +  \underbrace{ \cos \psi \hat{\boldsymbol{\varphi}}}_{\text{Azimuthal}} )
\text{,}
\end{split}
\label{eqn:inc_field}
\end{align}
where $w_0$ is the beam waist, $\psi$ is the spiral polarization angle, $\rho$ and $\varphi$ are the radial and axial cylindrical coordinates. Focusing the beam in Eqn.~\ref{eqn:inc_field} yields the following fields in the focal plane ($z=0$) near the optical axis~\cite{novotny_principles_2012}:
\begin{align}
\begin{split}
\label{eqn:focal_fields}
&\mathbf{E}^\text{foc}(x,y) \approx  \frac{1}{ \sqrt{\delta^2+1}} \left( \delta \underbrace{ ( x \hat{\boldsymbol {x}}   +y  \hat{\boldsymbol {y}} +  2\imath k^{-1} \hat{\boldsymbol {z}} )}_{\text{Radial}}  +  \underbrace{\left(x  \hat{\boldsymbol {y}}   -y  \hat{\boldsymbol {x}} \right) }_{\text{Azimuthal}} \right),\\
&\mathbf{H}^\text{foc}(x,y) \approx  \frac{\eta^{-1}}{ \sqrt{\delta^2+1}} \left(\delta \underbrace{\left(x  \hat{\boldsymbol {y}}   -y  \hat{\boldsymbol {x}} \right) }_{\text{Radial}}   - \underbrace{ ( x \hat{\boldsymbol {x}}   +y  \hat{\boldsymbol {y}} +  2\imath k^{-1} \hat{\boldsymbol {z}} )}_{\text{Azimuthal}} \right)
\text{,}
\end{split}
\end{align}
where $\delta=\tan \psi$ such that the values $\delta=0$ and $\delta \rightarrow \infty$ correspond to input azimuthal and radial polarizations, respectively. 
For the longitudinal fields, we used the paraxial approximation with a first order correction~\cite{Lax_parax,Bek_parax}. By substituting the focal fields (Eqns.~\ref{eqn:focal_fields}) into the expressions for directivity $D_x$ (Eqn.~\ref{eqn:dir2_full} and~\ref{eqn:dir2}) we get:
\begin{align}
\begin{split}
&D_x=  \frac{   \frac{4 }{k}  [ \delta (\delta y + x)  + \beta  ( y -\delta x)  ]  }{  (2 \delta/k)^2+ (\delta y +x )^2 +   \beta(2/k)^2+ \beta (y-\delta x)^2  },\\
&D_x(\beta=0)=  \frac{   \frac{4 \delta}{k}   (\delta y + x)   }{  (2 \delta/k)^2+ (\delta y +x )^2   }
\label{eqn:dir3_full} 
\text{,}
\end{split}
\end{align}
which leads to the following sensitivities $\Upsilon_x$ and angles $\theta_x$, as defined in Eqns.~\ref{eqn:sens1}:
\begin{align}
\begin{split}
&\Upsilon_x= \frac{k  \sqrt{  (\delta^2+\beta^2)(1+\delta^2)  }  }{ \delta^2+\beta},\\
&\theta_x = \tan^{-1}  \left(  \delta^2 + \beta ,\delta-\beta \delta \right),\\
&\Upsilon_x(\beta=0)= \frac{k}{ |\delta| }\sqrt{1+\delta^2} \propto |\delta|^{-1},\\
&\theta_x(\beta=0) = \tan^{-1}(\delta^2,\delta) =\psi + \pi/2 \left(1- \frac{\psi}{|\psi|} \right) 
\text{.}
\end{split}
\label{eqn:sens}
\end{align}
Eqns.~\ref{eqn:sens} are the main theoretical results of our manuscript. Specifically, they show that small values of $|\delta|  \approx |\psi| \ll 1$, which correspond to a mostly azimuthally polarized incident beam, significantly enhance the displacement sensitivity $\Upsilon_x$ for electric-dipole scatterers ($\beta=0$).

	\section{Experiment}
\begin{figure}
  \includegraphics[width=0.48\textwidth]{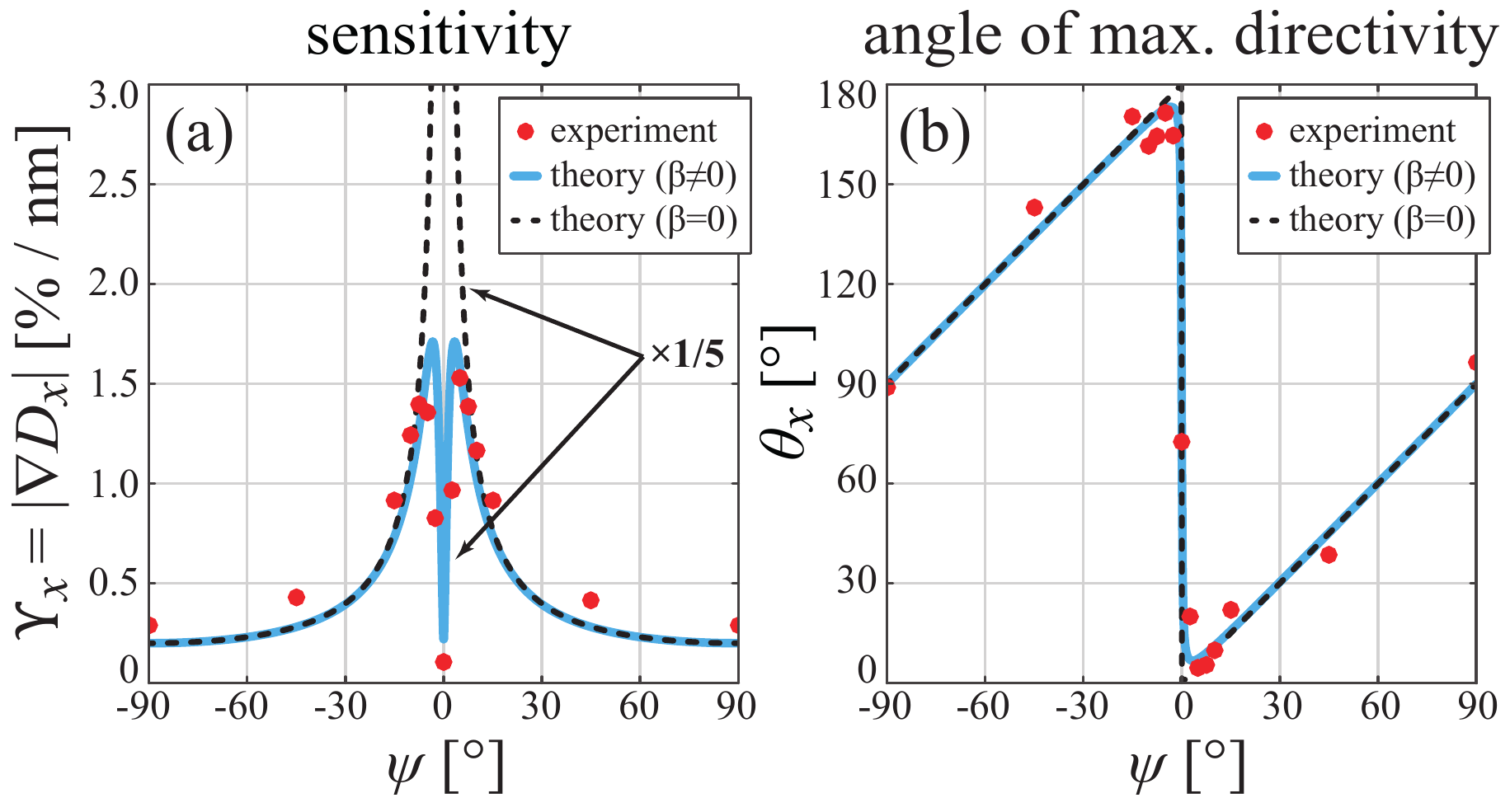}
  \caption{$(\boldsymbol{a})$ Sensitivity $\Upsilon_x$ as a function of the angle of spiral polarization $\psi$. The theoretical curves are divided by a factor of 5 for comparison. $(\boldsymbol{b})$ The corresponding angles $\theta_x$. The markers, solid blues lines and dashed black lines correspond to the experimental results, theoretical model in Eqns.~\ref{eqn:sens} for $\beta \neq 0$ and $\beta=0$, respectively.}
  \label{fig:sens}
\end{figure}

For an experimental demonstration, we use a spherical gold nanoparticle~\cite{Johnson} of diameter $d=120\,$nm, drop-casted onto a glass substrate with a refractive index of $n=1.52$. At a wavelengths of $\lambda=630\,$nm, the nanoparticle is approximately an electric dipole scatterer~\cite{Bohren_Absorption_1983} with $\beta=0.0034$. Fig.~\ref{fig:setup}$\boldsymbol{a}$ schematically depicts our experimental system~\cite{neugebauer_polarization_2014, neugebauer_magnetic_2017}. We start by converting an incident linearly polarized Gaussian beam into a spirally-polarized cylindrical vector beam using a q-plate of charge $1/2$~\cite{marrucci_optical_2006}. The relative angle between the axis of the q-plate and the polarization of the Gaussian beam defines the polarization angle $\psi$ of the spiral polarization beam in Fig.~\ref{fig:setup}$\boldsymbol{b}$. To ensure singlemodedness, we spatially filter the polarization tailored beam~\cite{karimi_hypergeometric-gaussian_2007}, which is subsequently focused by MO$_1$ with the numerical aperture of NA$_1=0.9$. The focal plane ($z=0$) coincides with the boundary between air ($z<0$) and the glass substrate ($z>0$). A high-precision 3D piezo actuator, attached to the glass substrate translates the nanoparticle through the focal volume of the beam. The confocally aligned MO$_2$ with NA$_2=1.4$ collects and collimates the transmitted and the scattered light, while the angular region of $\mathrm{NA}>\mathrm{NA}_1$ contains only scattered light. The collimated beam passes through a liquid crystal variable retarder set to $\lambda/4$ retardation and oriented at $45\degree$ with respect to the $x$-axis, which converts left- and right-hand circularly polarized components (LCP and RCP) to linearly polarized $x$ and $y$ components, respectively, preserving their intensity profiles $\mathrm{I}^{f}_{lcp}$ and $\mathrm{I}^{f}_{rcp}$. The subsequent polarizing beam splitter (PBS) separates the linear $x$ and $y$ polarizations. We simultaneously image the resulting projections onto a CMOS camera to visualize the intensity profiles of LCP and RCP components ($\mathrm{I}^{f}_{lcp}$ and $\mathrm{I}^{f}_{rcp}$) in the BFP of the collecting MO.\\%
\begin{figure}
  \includegraphics[width=0.48\textwidth]{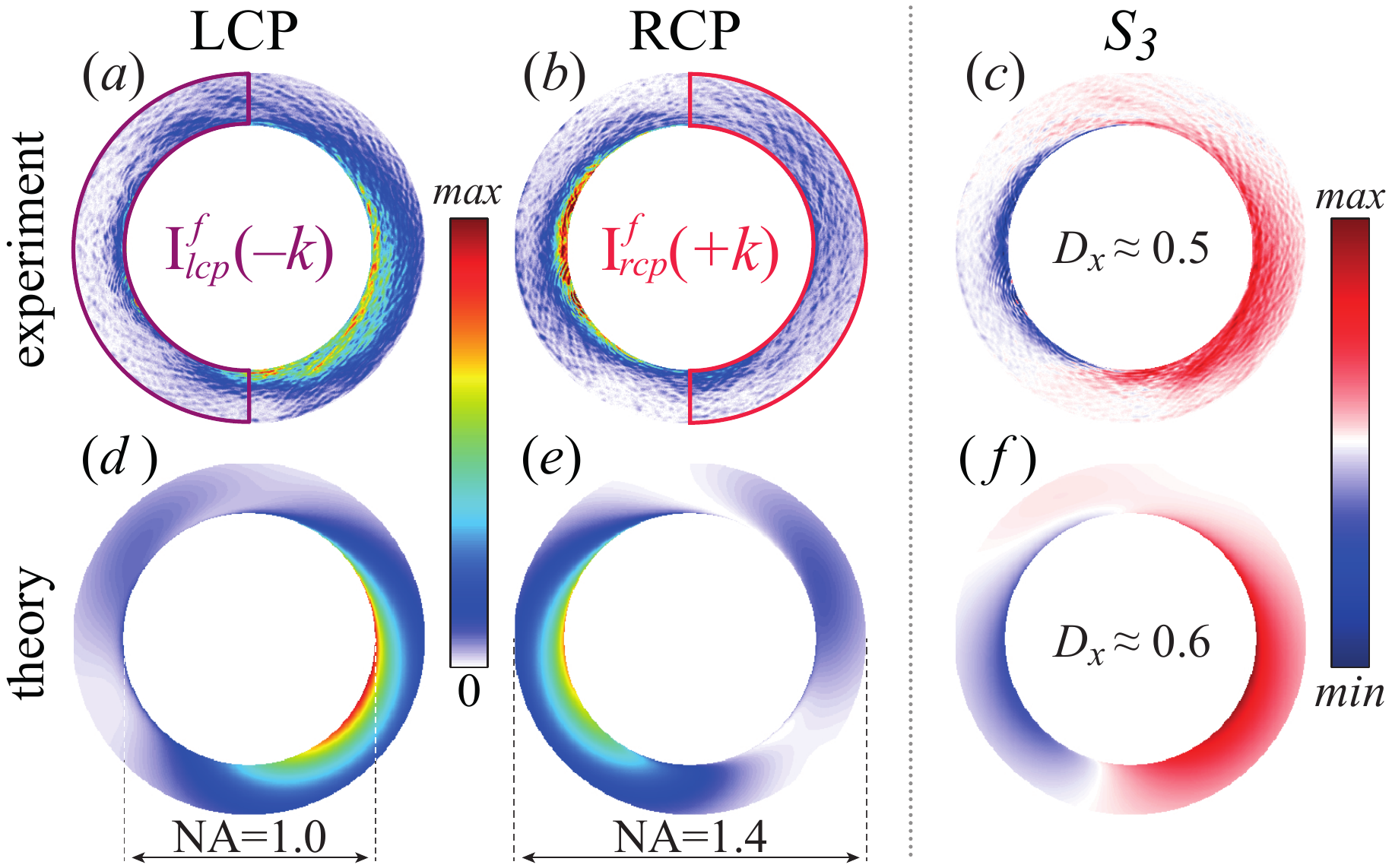}
  \caption{$(\boldsymbol{a})$, $(\boldsymbol{b})$ Experimentally and $(\boldsymbol{d})$, $(\boldsymbol{e})$ theoretically obtained (see Appendix A) left- and right-hand circularly polarized projections (LCP and RCP) $\mathrm{I}^{f}_{lcp}$ and $\mathrm{I}^{f}_{rcp}$ of the scattered light in the back focal plane of the second microscope objective for the position $(x,y)=(50\,\mathrm{nm},0)$ of the nanoparticle in the focal plane. $(\boldsymbol{c})$ and $(\boldsymbol{f})$ show a strong spin-segmentation in the associated third Stokes parameter $S_3 =\mathrm{I}^{f}_{lcp} - \mathrm{I}^{f}_{rcp}$. The brackets in $(\boldsymbol{a})$ and $(\boldsymbol{b})$ show the integration regions used on the experimental back focal plane images to calculate $D_x$ from Eqn.~\ref{eqn:dir1}.}
  \label{fig:bfp}
\end{figure}%
\begin{figure}
  \includegraphics[width=0.48\textwidth]{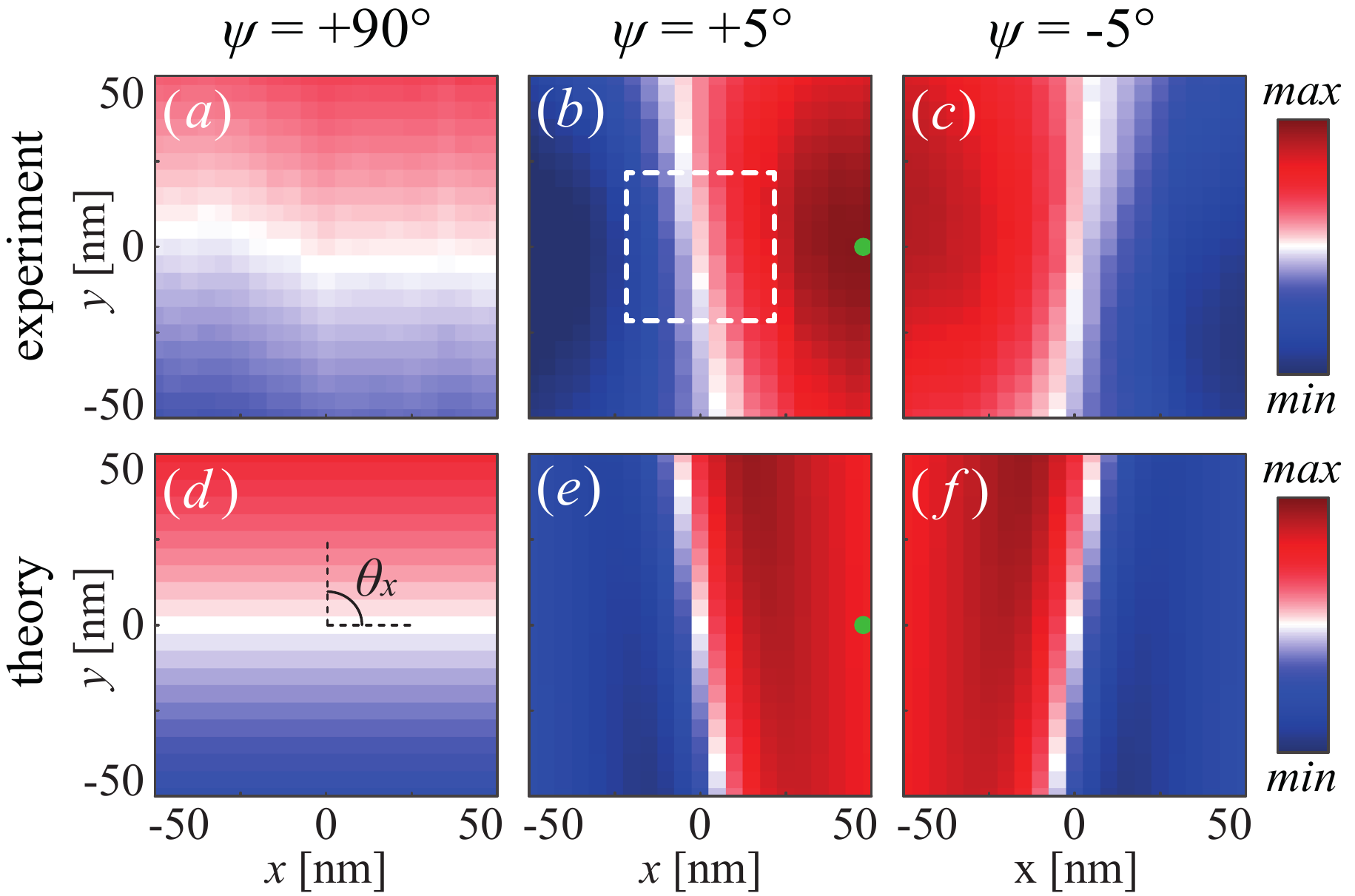}
  \caption{The experimental $(\boldsymbol{a})$-$(\boldsymbol{c})$ and theoretical $(\boldsymbol{d})$-$(\boldsymbol{f})$ values of the directivity $D_x(x,y)$ in far-field spin-segmentation with respect to $k_x$ for $\psi=+90 \degree ,\,+5\degree,\,-5\degree$. The green dots in $(\boldsymbol{b})$ and $(\boldsymbol{e})$ show the position $(x,y)=(50\,\mathrm{nm},0)$ in the focal plane. The dashed square in $(\boldsymbol{b})$ shows the area used to calculate sensitivity $\Upsilon_x$ and its corresponding angle $\theta_x$. }
  \label{fig:dir}
\end{figure}\\%
First, we translate the nanoparticle through the focal plane on a 2D grid of $50\,$nm~$\times~50\,$nm with a $5\,$nm step size for a variety of angles $\psi$ and record the polarization projections $\mathrm{I}^{f}_{lcp}$ and $\mathrm{I}^{f}_{rcp}$ at each point $(x,y)$ of the grid. In Fig.~\ref{fig:bfp}, we plot the individual experimental and theoretical BFP images (see Appendix A) obtained for $\psi= 5 \degree$. The top and bottom row in Fig.~\ref{fig:bfp} correspond to the position $(x,y)=(50\,\mathrm{nm},0)$ of the nanoparticle in the focal plane. There is a visible directivity in both LCP ($\mathrm{I}^{f}_{lcp}$) and RCP ($\mathrm{I}^{f}_{rcp}$) components of the scattered light, resulting in strong spin-segmentation in the associated third Stokes parameter $S_3 =\mathrm{I}^{f}_{lcp} - \mathrm{I}^{f}_{rcp}$, as shown in Fig.~\ref{fig:bfp}$\boldsymbol{c}$ and~\ref{fig:bfp}$\boldsymbol{f}$. Overall, we see an excellent qualitative correspondence between the experimental and theoretical data.
In order to increase the signal-to-noise ratio (see also Appendix B) and to calculate the experimental values of $D_x$ from the experimental BFP images, we integrate the intensities $\mathrm{I}^{f}_{lcp}$ and $\mathrm{I}^{f}_{rcp}$ over finite angular regions, which are marked in Fig.~\ref{fig:bfp}$\boldsymbol{a}$ and~\ref{fig:bfp}$\boldsymbol{b}$. We use the obtained values $\mathrm{I}^{f}_{lcp}(\pm k)$ and $\mathrm{I}^{f}_{rcp}(\pm k)$ to calculate $D_x$ in Eqn.~\ref{eqn:dir1}.\\%
In Fig.~\ref{fig:dir} we show the resulting experimental and theoretical (Eqn.~\ref{eqn:dir3_full}, $\beta\neq0$) plots of $D_x(x,y)$ for $\psi=90\degree, \, + 5\degree \, - 5\degree$. As expected from Eqn.~\ref{eqn:sens}, for radially polarized excitation ($\psi=90\degree$) the directivity in  far-field spin-segmentation with respect to $k_x$ appears for displacements of the nanoparticle along the $y$-axis ($\theta_x=\pi/2$)~\cite{rodriguez-herrera_optical_2010,rodriguez-herrera_far-field_2010,Nechayev_Weak_2018,Neugebauer_Weak_2019}, as shown in Fig.~\ref{fig:dir}$\boldsymbol{a}$ and~\ref{fig:dir}$\boldsymbol{d}$. We also notice in the second and third column of Fig.~\ref{fig:dir} a faster rise of $D_x$ for $\psi=\pm 5 \degree$, as compared to $\psi=90 \degree$ in the first column. The change of sign of $D_x$ and $\theta_x$ for $\psi=\pm 5 \degree$ is in agreement with Eqns.~\ref{eqn:sens}. The green dots in Fig.~\ref{fig:dir}$\boldsymbol{b}$ and~\ref{fig:dir}$\boldsymbol{e}$ correspond to the position $(x,y)=(50\,\mathrm{nm},0)$ of the nanoparticle in the focal plane for which we have plotted the BFP images in Fig.~\ref{fig:bfp}.\\%
Finally, to define the experimentally achieved sensitivities $\Upsilon_x$ and angles $\theta_x$, we fit the experimental plots of $D_x$ with planes in regions marked by a dashed square in Fig.~\ref{fig:dir}$\boldsymbol{b}$, such that:
\begin{align}
\begin{split}
&D_x= Ax +By + C,\\
&\Upsilon_x= \sqrt{A^2+B^2},\\
&\theta_x=\tan^{-1}(B,A)
\text{.}
\end{split}
\label{eqn:2dfit}
\end{align}
In Fig.~\ref{fig:sens}$\boldsymbol{a}$ we plot the theoretical (Eqns.~\ref{eqn:sens}) and experimental (solid blue line and red markers, respectively) values of $\Upsilon_x$~\cite{symmetry}, showing that we correctly resolve the characteristic shape of $\Upsilon_x$. The fast decay of $\Upsilon_x$ for $\psi \rightarrow 0 \degree$, following two sharp maxima around $\psi  \approx \pm 3.3 \degree$, is the result of the selective excitation of a magnetic dipole moment and simultaneous suppression of the electric dipole moment in the nanoantenna for azimuthally polarized incident beam~\cite{Wozniak_Selective_2015}. We also show the theoretical curve obtained from Eqns.~\ref{eqn:sens} for $\beta=0$ as a black dashed line in Fig.~\ref{fig:sens}$\boldsymbol{a}$, which approaches infinity for $\psi \rightarrow 0 \degree$. We obtain a maximal experimental value of $\Upsilon_x$ of $\approx 1.5\,\%\,\mathrm{nm}^{-1}$ for $\psi = \pm 5 \degree$, which is approximately 5 times smaller than in theory. This discrepancy mostly originates from our very simplified free-space model that does not account for the glass substrate~\cite{novotny_principles_2012,substrate_halas,substrate_bianisotropy} (see Appendix A), for the exact form of the focal fields~\cite{novotny_principles_2012}, for the integration regions in the BFP and for the plane fitting of $D_x$. A theoretical point-dipole model that includes these effects reduces the discrepancy already to a factor of 2. Additionally, an ideal point-dipole model still neglects the experimental imperfections, e.g., the extinction ratios of polarizers, phase shifts introduced by the optical elements, actual shape of the nanoparticle and, most importantly, the beam preparation and experimentally achieved focal fields.\\%
In Fig.~\ref{fig:sens}$\boldsymbol{b}$ we plot the theoretical (Eqns.~\ref{eqn:sens}) and experimental values of $\theta_x$~\cite{symmetry}. The simplified theory in Eqns.~\ref{eqn:sens} shows good correspondence to the experimental values of $\theta_x$ for both $\beta \neq 0$ and $\beta = 0 $. Since the value of $\theta_x$ strongly correlates with the angle of the spiral polarization $\psi$ (Eqns.~\ref{eqn:sens}), it also serves as a verification of the form of the incident beam in Eqn.~\ref{eqn:inc_field} and the quality of preparation of the focal fields in Eqns.~\ref{eqn:focal_fields}.\\%
\\
\section{Discussion and Conclusion}
In summary, we have considered a sub-wavelength localization and displacement sensing scheme that is based on position-dependent scattering response of optical nanoantennas excited with structured light beams. We have shown a characteristic feature of this and similar schemes, that the far-field sensitivity to lateral displacement of nanoantennas depends on the spatial derivatives of the excitation field. Experimentally, we have investigated the specific case of far-field spin-segmentation upon scattering of tightly focused beams bearing transverse spin by an electric-dipole nanoantenna. By designing focal field gradients, we have strongly enhanced the sensitivity of this segmentation to lateral displacements of the nanoantenna within the focal plane of the beam. The maximal achieved sensitivity $\Upsilon_x$ of $ \approx 1.5\,\%\,\mathrm{nm}^{-1}$ is comparable to previous experimental records of $ \approx 1\,\%\,\mathrm{nm}^{-1}$~\cite{roy_radially_2015,neugebauer_polarization-controlled_2016}, $ \approx 2.5\,\%\,\mathrm{nm}^{-1}$~\cite{Bag2018} and $ \approx 5\,\%\,\mathrm{nm}^{-1}$~\cite{Neugebauer_Weak_2019}. In principle, $\Upsilon_x$ has a potential for further improvement with better optical components in the current experimental scheme. However, Fig.~\ref{fig:sens}$\boldsymbol{a}$ clearly shows that the limiting factor in our experiment is the excitation of a non-negligible magnetic dipole moment, i.e., the validity of the electric-dipole approximation of our scatterer. Recently, intelligent designs of core-shell nanoparticles have been suggested aiming to enhance or suppress specific multipolar contributions to the scattering response. Specifically, in~\cite{ideal_magnetic}, the authors theoretically predict an ideal magnetic dipole scattering. Applying our scheme to such a scatterer to enhance the magnetic counterpart of the giant spin-Hall effect of light may lead to further enhanced sensitivity that approaches the dashed curve in Fig.~\ref{fig:sens}$\boldsymbol{a}$.\\
Importantly, the results presented in sections II\&III are by no means limited to focal fields featuring transverse spin density, to observation of circular polarization effects or to dipolar scatterers. The sensitivity of any localization or displacement-sensing scheme relying on spatial inhomogeneity of the excitation field depends on the spatial derivatives of this field. For instance, one can chose a $\pm \pi/2$ dephased superposition of radial and azimuthal polarization in Eqn.~\ref{eqn:inc_field} to observe far-field segmentation of the first $S_1$ and second $S_2$ Stokes parameters, as directly follows from the derivations in sections II.\\
Finally, we would like to note a similarity between our approach to enhance the displacement sensitivity and the quantum weak measurements inspired~\cite{aharonov_how_1988,duck_sense_1989} polarization filtering techniques, typically employed to enhance the visibility of beam shift phenomena~\cite{Zeld_observed_1994,hosten_observation_2008,gorodetski_weak_2012,Magana-Loaiza2014}. In the context of classical optics, a ``weak measurement'' with pre- and post-selected state vectors consist of performing a measurement in which polarization weakly couples to other degrees of freedom using two almost orthogonal input and output polarizations states~\cite{weak_phase2_2010,weak_phase_2013,weak_pulse_2014,weak_or_2016,weak_birefr_bl_2016}. Only recently these techniques have expanded to the realm of nano-optics and nano-photonics~\cite{Nechayev_Weak_2018,Neugebauer_Weak_2019,Araneda2019}, for which an analogy was put forward~\cite{Nechayev_Weak_2018}. Also in our case, a superposition of azimuthal polarization with a small fraction of radial polarization and the selective response of the electric-dipole nanoantenna to the focal electric field created mostly by the radial component near the optical axis can be regarded as pre- and post-selection, respectively.\\
In conclusion, we believe that precise sculpturing of electromagnetic field gradients with intelligent design of nanoantennas' scattering response open new avenues for drastically increased displacement sensitivity. These capabilities may have far-reaching implications in nanophotonics, microscopy and beyond.
\begin{acknowledgments}
The authors acknowledge fruitful discussions with Gerd Leuchs.
\end{acknowledgments}
\appendix
\section{Dipole nanoantenna on a substrate}
For a more complete theoretical description of the experiment, we use an exact calculation of the focal fields that includes reflections at the interface~\cite{novotny_principles_2012} with the experimental beam waist of $w_0=1.6\,\mathrm{mm}$ and focal distances of the MOs of $f=2\,\mathrm{mm}$. Next, we account for the substrate-induced bi-anisotropy in the dipole polarizabilites~\cite{substrate_halas,substrate_bianisotropy}. The excited dipole moments are 
\begin{align}
\begin{split}
&\mathbf{p}\left(x,y\right)=\hat{\alpha}_{ee} \varepsilon_0 \mathbf{E}^\text{foc}\left(x,y\right)+\hat{\alpha} _{em}\mathbf{H}^\text{foc}\left(x,y\right),\\
&\mathbf{m}\left(x,y\right)=\hat{\alpha}_{mm} \mathbf{H}^\text{foc}\left(x,y\right)+\hat{\alpha}_{me} \mathbf{E}^\text{foc}\left(x,y\right)
\text{,}
\end{split}
\label{eqn:binais}
\end{align}
with values of $\hat{\alpha}_{ij}$ calculated according to Ref.~\cite{substrate_bianisotropy} using Mie coefficients~\cite{Bohren_Absorption_1983} of a spherical gold nanoparticle~\cite{Johnson} of diameter $d=120\,$nm located at a distance of $d/2$ above a glass substrate with the refractive index of $n=1.52$.\\
We assume that the point-dipole is positioned in air (half-space $z<0$) above the glass substrate (half-space with $z>0$) at $(x,y,-d/2)$. The far-field emission of $\mathbf{p}(x,y)$ and $\mathbf{m}(x,y)$ into the hemisphere $z>0$, as observed in the BFP of MO$_2$, according to Ref.~\cite{novotny_principles_2012} is given in $(p,s)^\top$ basis by:
\begin{align}
\begin{split}
\label{eqn:farfield_exact}
 &\mathbf{E}^{f}_{\mathbf{p}}(k_{x},k_{y})
 = CD(x,y)
\left[\begin{matrix}
\frac{k_{x}k_{z}}{k_{\bot}k}t_{p} &\frac{k_{y}k_{z}}{k_{\bot}k}t_{p}&-\frac{k_{\bot}}{k}t_{p}\\
-\frac{k_{y}}{k_{\bot}}t_{s} &\frac{k_{x}}{k_{\bot}}t_{s}&0
\end{matrix}\right]
\mathbf{p}(x,y),\\
&\mathbf{E}^{f}_{\mathbf{m}}(k_{x},k_{y})
 = CD(x,y)
\left[\begin{matrix}
-\frac{k_{y}}{k_{\bot}}t_{p} &\frac{k_{x}}{k_{\bot}}t_{p}&0\\
-\frac{k_{x}k_{z}}{k_{\bot}k}t_{s} &-\frac{k_{y}k_{z}}{k_{\bot}k}t_{s}&\frac{k_{\bot}}{k}t_{s}
\end{matrix}\right]
\frac{\mathbf{m}(x,y)}{c},\\
&D(x,y)=\exp{\left[\imath \left(k_x x+k_y y + k_z d \right) \right]} \times \frac{ \exp{\left[\imath nkf \right]}}{f},\\
&C= \frac{k^2}{4 \pi \varepsilon_0 } \times \frac{ \sqrt{ k^{2}n^{2}-k_{\bot}^{2} }  }{k_z} \times  \frac{ \sqrt{ nk }  }{( k^{2}n^{2}-k_{\bot}^{2})^{1/4}}
\text{.}
\end{split}
\end{align}
Here, $t_{p}\left(k_{\bot}\right)$ and $t_{s}\left(k_{\bot}\right)$ are the Fresnel transmission coefficients, \mbox{$k_{\bot}=\left(k_{x}^2+k_{y}^2\right)^{1/2}\leq nk$} is the transverse wavenumber and \mbox{$k_{z}=\left(k^{2}-k_{\bot}^{2}\right)^{1/2}$} is the longitudinal wavenumber with $\Im\left[k_{z}\right]\geq0$. In the BFP of MO$_2$ only the scattered light appears in the angular region \mbox{$\mathrm{NA}_1 < k_{\bot} \leq \mathrm{NA}_2$}. We used Eqns.~\ref{eqn:binais}-\ref{eqn:farfield_exact} to calculate the theoretical BFP images in Fig.~\ref{fig:bfp}$\boldsymbol{d}$-\ref{fig:bfp}$\boldsymbol{f}$. Additionally, we have applied the same procedure of data analysis and plane-fitting on the theoretical BFP images as on the experimental ones, which significantly improved our prediction in Fig.~\ref{fig:sens}$\boldsymbol{a}$ (not shown here). However, the qualitative behavior of the improved model matches the simplified free-space model provided by Eqns.~\ref{eqn:sens}.%
\section{Signal-to-noise ratio}
In section IV we have mentioned that our approach to enhance the displacement sensitivity resembles quantum weak measurement inspired polarization filtering techniques~\cite{aharonov_how_1988,duck_sense_1989,hosten_observation_2008,gorodetski_weak_2012,Magana-Loaiza2014}. There is an ongoing debate on whether these techniques allow for enhancing signal-to-noise (SNR) ratio or yield more Fisher information. We refer the readers to~\cite{Escher2011,wm_snr1,wm_snr2,wm_snr3,wm_snr4,wm_snr5,wm_snr6,wm_snr7} for a more elaborate discussion. However, it is instructive to find out whether the drastic increase of sensitivity $\Upsilon_x$ for an ideal electric-dipole nanoantenna ($\beta=0$) shown by the dashed black line in Fig.~\ref{fig:sens} influences the SNR of the measurement. We assume a shot noise limited scenario, in which the noise is given by the square root of the total intensity $\Delta I = \sqrt{I} = \sqrt{S_0}$, while the signal corresponds to the third Stokes parameter $\left|S_3\right|$. The SNR is therefore given by:
\begin{align}
\begin{split}
\text{SNR}(x,y,\delta)=\frac{\left|S_3\right|}{\sqrt{S_0}}= P \frac{ \left|2   \Im  \left\{ \mathrm{E}_z^\text{foc} \left( \mathrm{E}_y^\text{foc}  \right) ^*  \right\}  \right|  }{ \sqrt{ | \mathrm{E}_y^\text{foc} |^2+| \mathrm{E}_z^\text{foc} |^2} }
\text{,}
\end{split}
\label{eqn:snr1}
\end{align}
where we have combined all constants that do not depend on $(x,y,\delta)$ into the proportionality constant $P$. Using the focal field in Eqns.~\ref{eqn:focal_fields} yields:
\begin{align}
\begin{split}
\text{SNR}(x,y,\delta)=\frac{P}{\sqrt{\delta^2+1}} \frac{  \left| \frac{4 \delta}{k}   (\delta y + x)  \right| }{ \sqrt{  (2 \delta/k)^2+ (\delta y +x )^2  } }
\text{.}
\end{split}
\label{eqn:snr2}
\end{align}
We now have to examine Eqn.~\ref{eqn:snr2} for small displacements $(x,y)$ in the limit of radial ($\delta \rightarrow  \infty$) and almost azimuthal ($\delta \approx 0$) input polarizations. Since the maximal directivity $D_x(\beta=0)=1$ in Eqn.~\ref{eqn:dir3_full} is achieved for any $(x,y)$ on the line of $x+\delta y=2\delta/k$, e.g., for $(x,y)=(0,2/k)$ or $(x,y)=(2\delta/k,0)$, we consider displacements such that $0<\left|x\right|=\left|y\right|=r \ll \delta/k$. Under these assumptions, for radially polarized ($\delta \rightarrow  \infty$) and almost azimuthally polarized ($\delta \approx 0$) input beams we get:
\begin{align}
\begin{split}
&\text{SNR}^{\text{azi}}(r,\delta \approx 0)=\text{SNR}^{\text{rad}}(r,\delta \rightarrow \infty)=2Pr
\text{.}
\end{split}
\label{eqn:snr3}
\end{align}
Consequently, in our case, the drastic increase of sensitivity to small displacements does not influence SNR in shot noise limited scenarios~\cite{Escher2011,wm_snr1,wm_snr2,wm_snr3,wm_snr4,wm_snr5,wm_snr6,wm_snr7}.
\bibliography{Manusc_weakm_local}

\begin{thebibliography}{60}
\bibitem [{sym()}]{symmetry}%
  \BibitemOpen
  \href@noop {} {}\bibinfo {note} {Owing to the cyllindrical symmetry of our
  system, we also calculate the similarly defined values of $\Upsilon_y$ and
  $\theta_y$ and plot in Fig.~\ref{fig:sens} the experimental averages
  $(\Upsilon_x+\Upsilon_y)/2$ and $(\theta_x+\theta_y)/2$.}\BibitemShut {Stop}%
\end{thebibliography}
\end{document}